\documentclass{elsart}
\usepackage{epsfig}

\begin{document}
\begin{frontmatter}
\title{\large\bf\boldmath Measurements of Cabibbo Suppressed Hadronic Decay Fractions
       of Charmed $D^0$ and $D^+$ Mesons}

\begin{small}
\begin{center}
\vspace{0.2cm}

M.~Ablikim$^{1}$, J.~Z.~Bai$^{1}$, Y.~Ban$^{11}$,
J.~G.~Bian$^{1}$, X.~Cai$^{1}$, J.~F.~Chang$^{1}$,
H.~F.~Chen$^{16}$, H.~S.~Chen$^{1}$, H.~X.~Chen$^{1}$,
J.~C.~Chen$^{1}$
% ...
\\(BES Collaboration)\\
\vspace{0.2cm}
\label{att}
$^{1}$ Institute of High Energy Physics, Beijing 100049, People's Republic of China\\
$^{2}$ China Center for Advanced Science and Technology (CCAST),
Beijing 100080, People's Republic of China\\
$^{3}$ Guangxi Normal University, Guilin 541004, People's Republic of China\\
$^{4}$ Guangxi University, Nanning 530004, People's Republic of China\\
$^{5}$ Henan Normal University, Xinxiang 453002, People's Republic of China\\
$^{6}$ Huazhong Normal University, Wuhan 430079, People's Republic of China\\
$^{7}$ Hunan University, Changsha 410082, People's Republic of China\\
$^{8}$ Liaoning University, Shenyang 110036, People's Republic of China\\
$^{9}$ Nanjing Normal University, Nanjing 210097, People's Republic of China\\
$^{10}$ Nankai University, Tianjin 300071, People's Republic of China\\
$^{11}$ Peking University, Beijing 100871, People's Republic of China\\
$^{12}$ Shandong University, Jinan 250100, People's Republic of China\\
$^{13}$ Shanghai Jiaotong University, Shanghai 200030, People's Republic of China\\
$^{14}$ Sichuan University, Chengdu 610064, People's Republic of China\\
$^{15}$ Tsinghua University, Beijing 100084, People's Republic of China\\
$^{16}$ University of Science and Technology of China, Hefei 230026, People's Republic of China\\
$^{17}$ Wuhan University, Wuhan 430072, People's Republic of China\\
$^{18}$ Zhejiang University, Hangzhou 310028, People's Republic of China\\
\vspace{0.4cm}

$^{a}$ Current address: Iowa State University, Ames, IA 50011-3160, USA.\\
$^{b}$ Current address: Purdue University, West Lafayette, IN 47907, USA.\\
$^{c}$ Current address: Cornell University, Ithaca, NY 14853, USA.\\
$^{d}$ Current address: Laboratoire de
l'Acc{\'e}l{\'e}ratearLin{\'e}aire,
F-91898 Orsay, France.\\
$^{e}$ Current address: DESY, D-22607, Hamburg, Germany.\\

\end{center}
\end{small}
\maketitle

\normalsize

\begin{abstract}

Using data collected with the BESII detector at $e^{+}e^{-}$
storage ring Beijing Electron Positron Collider, the measurements
of relative branching fractions for seven Cabibbo suppressed
hadronic weak decays $D^0 \rightarrow K^- K^+$, $\pi^+ \pi^-$,
$K^- K^+ \pi^+ \pi^-$ and $\pi^+ \pi^+ \pi^- \pi^-$, $D^+
\rightarrow \overline{K^0} K^+$, $K^- K^+ \pi^+$ and $\pi^- \pi^+
\pi^+$ are presented.

\vspace{3\parskip} \noindent{\it PACS:} 13.25.Ft, 14.40.Lb
\end{abstract}
\end{frontmatter}


\section{INTRODUCTION}
\label{secintro}

\begin{figure}[htbp]
  \centering
  \begin{minipage}[b]{0.42\linewidth}
  \centering
  \includegraphics[height=2.0cm,width=3.0cm]{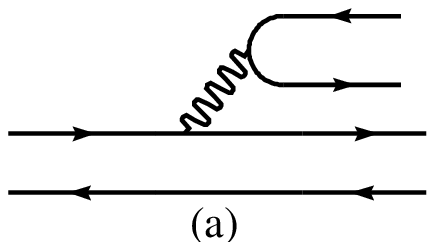}
  \end{minipage}
  \begin{minipage}[b]{0.42\linewidth}
  \centering
  \includegraphics[height=2.0cm,width=3.0cm]{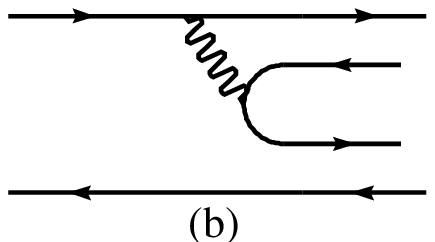}
  \end{minipage}
  \begin{minipage}[b]{0.42\linewidth}
  \centering
  \includegraphics[height=2.0cm,width=3.0cm]{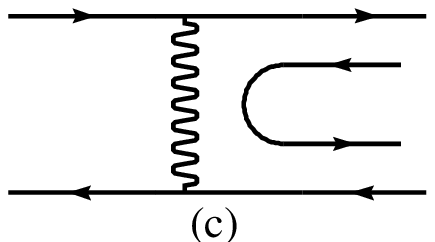}
  \end{minipage}
  \begin{minipage}[b]{0.42\linewidth}
  \centering
  \includegraphics[height=2.0cm,width=3.0cm]{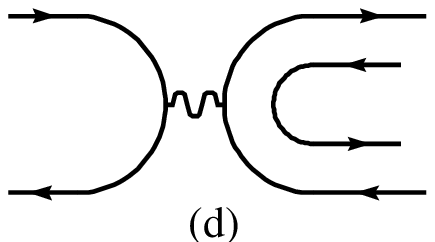}
  \end{minipage}
  \begin{minipage}[b]{0.42\linewidth}
  \centering
  \includegraphics[height=2.0cm,width=3.0cm]{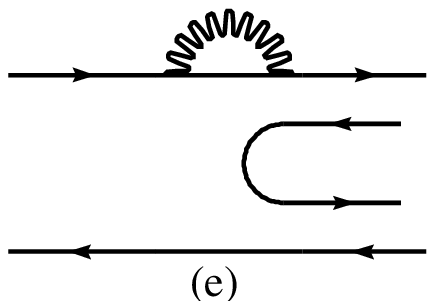}
  \end{minipage}
  \begin{minipage}[b]{0.42\linewidth}
  \centering
  \includegraphics[height=2.0cm,width=3.0cm]{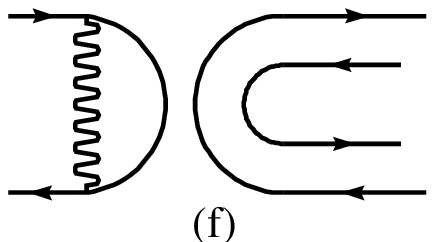}
  \end{minipage}
  \caption{
  Typical Feynman diagrams for Cabibbo suppressed decays of charm
  mesons.}
  \label{feynme}
\end{figure}

Hadronic decays of charmed mesons have been extensively studied.
Measurements of relative lifetimes and semileptonic branching
fractions for charmed mesons $D^+$ and $D^0$ suggest the presence
of nonleptonic processes which enhance the $D^0$ and suppress the
$D^+$ width, and lead to the conclusion that the
simple spectator model of charmed-meson decay is inadequate. As
shown in Figure \ref{feynme}, all weak decays of heavy mesons may
be described by six quark-diagrams: the external W-emission
diagrams (a), the internal W-emission diagrams (b), the W-exchange
diagram (c), the W-annihilation diagram (d), the horizontal W-loop
diagram (e), and the vertical W-loop diagram (f) \cite{sixfey}.
Thus, a further understanding of the $D$ decay mechanism, such as the
contributions of other quark-diagrams and final states interactions,
requires a systematic study of the hadronic decays.

At present, many experiments, such as, MARKII \cite{mk2}, MARKIII
\cite{mk3}, E691 \cite{e691}, E687 \cite{e687}, E791 \cite{e791},
FOCUS \cite{focus} and CLEO \cite{cleo}, have reported their
measurements of Cabibbo suppressed hadronic decay fractions of $D$
mesons. Our present measurements are based on a data sample of
integrated luminosity of $\sim 17.3 \mathrm{pb^{-1}}$ at
$\psi(3770)$ peak ($\sqrt{s}=3.773$ GeV) and $\sim
16.5\mathrm{pb^{-1}}$ for $\psi(3770)$ peak scan collected by
Beijing Spectrometer (BESII) detector at $e^{+}e^{-}$ storage ring
Beijing Electron Positron Collider (BEPC) \cite{bepc}. This paper
reports our measurements of Cabibbo suppressed relative branching
fractions for several hadronic decay modes of  charmed $D$ mesons
$D^0 \rightarrow K^- K^+$, $\pi^+ \pi^-$, $K^- K^+ \pi^+ \pi^-$,
$\pi^+ \pi^+ \pi^- \pi^-$, $D^+ \rightarrow \overline{K^0} K^+$,
$K^- K^+ \pi^+$ and $\pi^- \pi^+ \pi^+$ (through out this paper
the charge conjugate states are implicitly included).

\section{BESII DETECTOR}
\label{secbes}

The Beijing Spectrometer (BESII) is a conventional cylindrical
magnetic detector that is described in detail in Ref. \cite{bes2}.
A 12-layer Vertex Chamber (VC) surrounds the beryllium beam pipe
and provides trigger information, as well as coordinate
information. A forty-layer main drift chamber (MDC) located just
outside the VC yields precise measurements of charged particle
trajectories with a solid angle coverage over $85\%$ of $4\pi$; it
also provides ionization energy loss ($dE/dx$) measurements which
are used for particle identification.  Momentum resolution of
$1.7\%\sqrt{1+p^2}$ ($p$ in GeV/$c$) and $dE/dx$ resolution for
hadron tracks of $\sim8\%$ are obtained.  An array of 48
scintillation counters surrounding the MDC measures the time of
flight (TOF) of charged particles with a resolution of about 200
ps for hadrons. Outside the TOF counters, a 12 radiation length,
lead-gas barrel shower counter (BSC), operating in limited
streamer mode, measures the energies of electrons and photons over
$80\%$ of the total solid angle with an energy resolution of
$\sigma_{E}/E=0.22/\sqrt{E}$ ($E$ in GeV). Outside the solenoidal
coin, which provides a 0.4 T magnetic field over the tracking
volume, is an iron flux return that is instrumented with three
double-layer muon counters that identify muons with momentum
greater than 500 MeV$/c$.

In this analysis, a GEANT3 based Monte Carlo package (SIMBES
\cite{simbes}) with detailed consideration of the detector
performance (such as dead electronic channels) is used. The
consistency between data and Monte Carlo has been carefully
checked in many high purity physics channels, and the agreement is
reasonable.\par

\section{EVENT SELECTION}
\label{secevt}

Charged tracks are required to satisfy $\left| \cos \theta \right|
< 0.8$, where $\theta$ is the polar angle in the MDC, and have
good helix fit. The tracks that are not associated with $K^0_S$
reconstruction are required to be originate from interaction
point.
 Pions and kaons are identified by requiring the confidence
level of desired hypothesis using combined measurements of
time-of-flight \cite{tof} and energy loss in drift chamber to be
greater than 0.1\%. In addition, 
kaon
and pion candidates are further identified by requiring the
normalized weights, which is defined as
${CL}_{\alpha}/({CL}_{\pi}+{CL}_{K})$, where $\alpha$ denotes
desired particle, exceeding $50\%$. 
\par
$K^0_S$ candidates are detected through the decay of $K^0_S
\rightarrow \pi^+ \pi^-$. Each oppositely charged track pair is
assumed to be $\pi^+$ and $\pi^-$. The decay vertex of $K^0_S$ is
required to be 5mm far away from the beam axis.
The $\pi^+ \pi^-$ invariant mass is required to be within 0.020
GeV/$c^2$ of the $K^{0}_{S}$ nominal mass.\par Cabibbo suppressed
hadronic decay modes are expected at a lower rate ($\sim
\tan^2\theta_{C}$, where $\theta_{C}$ is ``Cabibbo angle'')
compared to relevant Cabibbo favoured modes, for which $\pi/K$
misidentification becomes significant. The unique energy of $D$
meson at the $\psi(3770)$ can be exploited to reduce explicitly
background due to incorrect particle assignment. A single particle
misidentification results in a reflection peak separated from the
beam energy. \par
\begin{figure}[htbp]
  \centering
  \includegraphics[height=10.0cm,width=8.0cm]{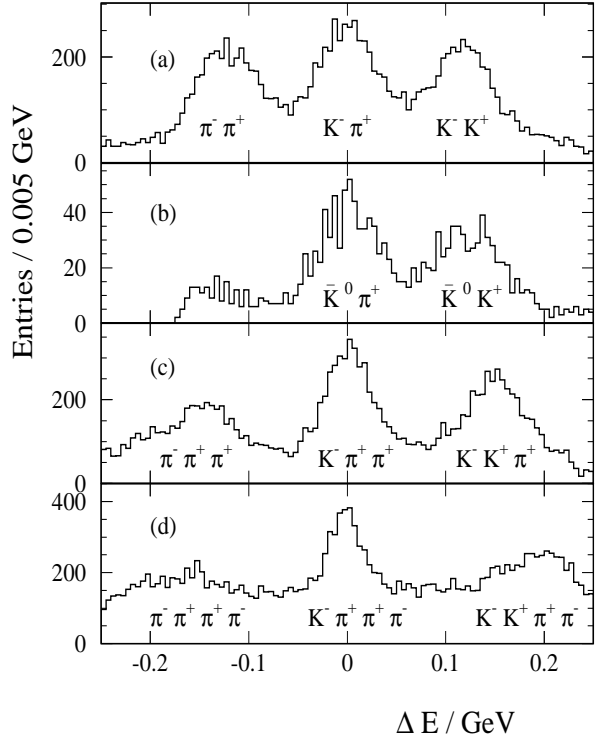}
\caption[]{ $\Delta E$ distribution and reflection for similar
Cabibbo suppressed modes by using the sample of Cabibbo favoured
mode (a) $D^0\rightarrow K^- \pi^+$ (b) $D^0\rightarrow
\overline{K^0} \pi^+$ (c) $D^+\rightarrow K^- \pi^+ \pi^+$ (d)
$D^0 \rightarrow K^{-} \pi^{+}\pi^{+}\pi^{-}$. The distributions
show the well-separated $\Delta E$ peaks.} \label{dele}
\end{figure}

 The distributions of energy difference($\Delta
E$) between measured energy of $D$ candidates ($E_{\mathrm{tag}}$)
and beam energy ($E_{\mathrm{b}}$) are shown in Figure \ref{dele}
for 4 Cabibbo allowed decay modes with correct $\pi / K$
assignments and the reflection $\Delta E$ distribution with a
single particle misidentification. $\Delta E$ of $D$ candidate is
required to be less than that of similar topological decay mode to
reject particle misidentification combination. $\Delta E$ is
further required to be within 50-100 MeV for different decay
channels. 

The pair production of $D\overline{D}$ at $\psi(3770)$ provides a
variable, which is defined as the beam-constrained
mass
$$
M_{\mathrm{bc}}=\sqrt{E^2_{\mathrm{beam}} - (\sum_{\mathrm{i}} p_{\mathrm{i}})^2}
$$
exploiting the fact that the total energy of all decay products
must sum to the beam energy. As the uncertainty in the beam energy
is much smaller than the uncertainty in the total reconstructed
energy of the decay tracks, this approach yields much improved
mass resolution compared to the invariant mass technique. \par
The QED processes, $\tau^{+}\tau^{-}$ pair productions and cosmic
backgrounds may contribute to $D$ tags. Both of them have a lower
multiplicity than that of $D\overline{D}$ decay, the requirement
of $\displaystyle N_{\mathrm{ch}}+N_{\mathrm{neu}}/2>3$ will
eliminate most of these backgrounds, where $N_{\mathrm{ch}}$ and
$N_{\mathrm{neu}}$ represent the total number of charged
tracks and neutral tracks respectively. At $\psi(3770)$, $D
\overline{D}$ are produced with the angular distribution $\sin^2
\theta_D$, where $\theta_{D}$ is the production angle of
$\psi(3770)\rightarrow D \overline{D}$, $\left|\cos
\theta_{D}\right| < 0.8$ is imposed to each $D$ tag to enhance signal
to background ratio.\par
\begin{figure}
  \centering
  \includegraphics[width=6.5cm]{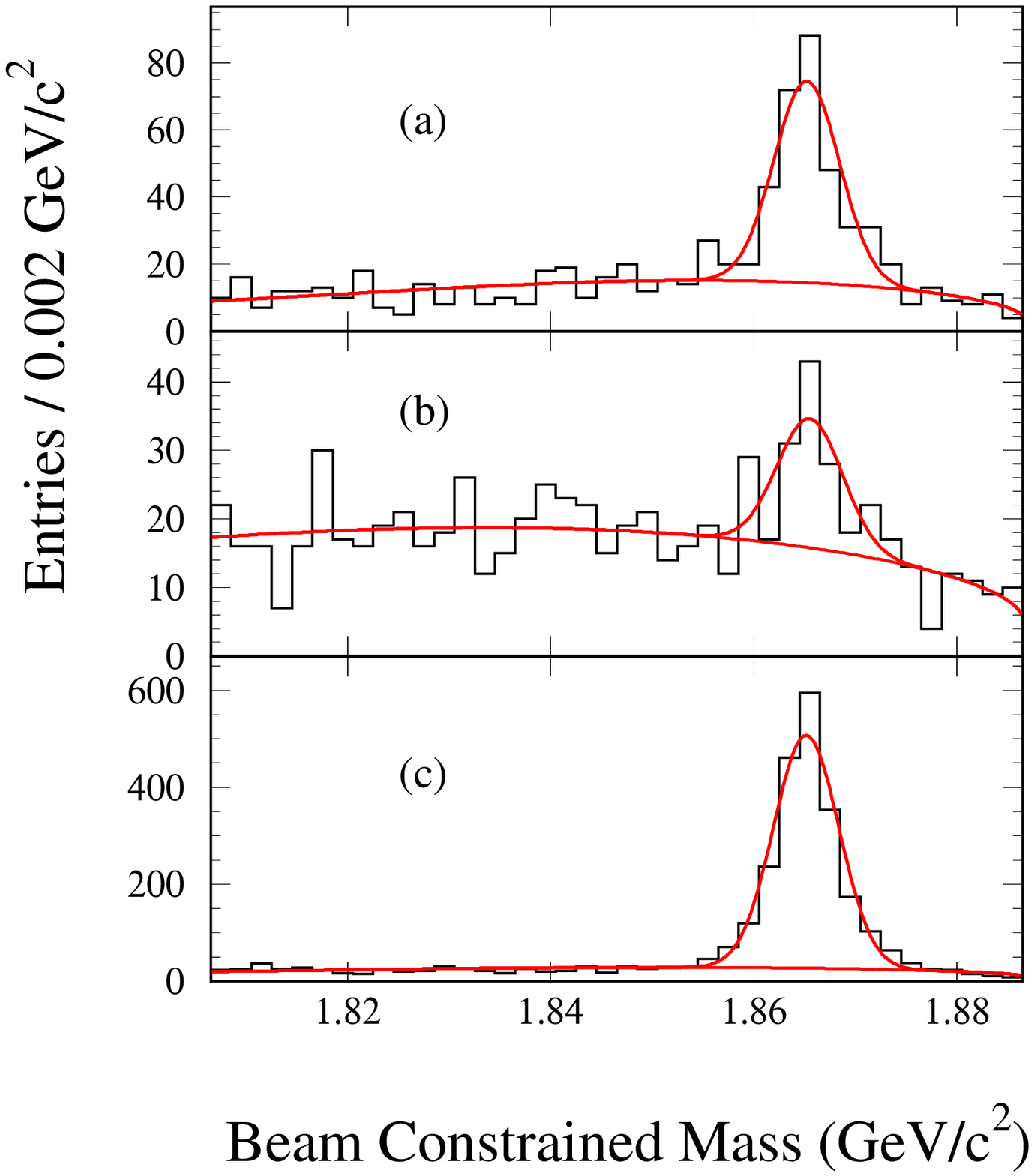}
  \includegraphics[width=6.5cm]{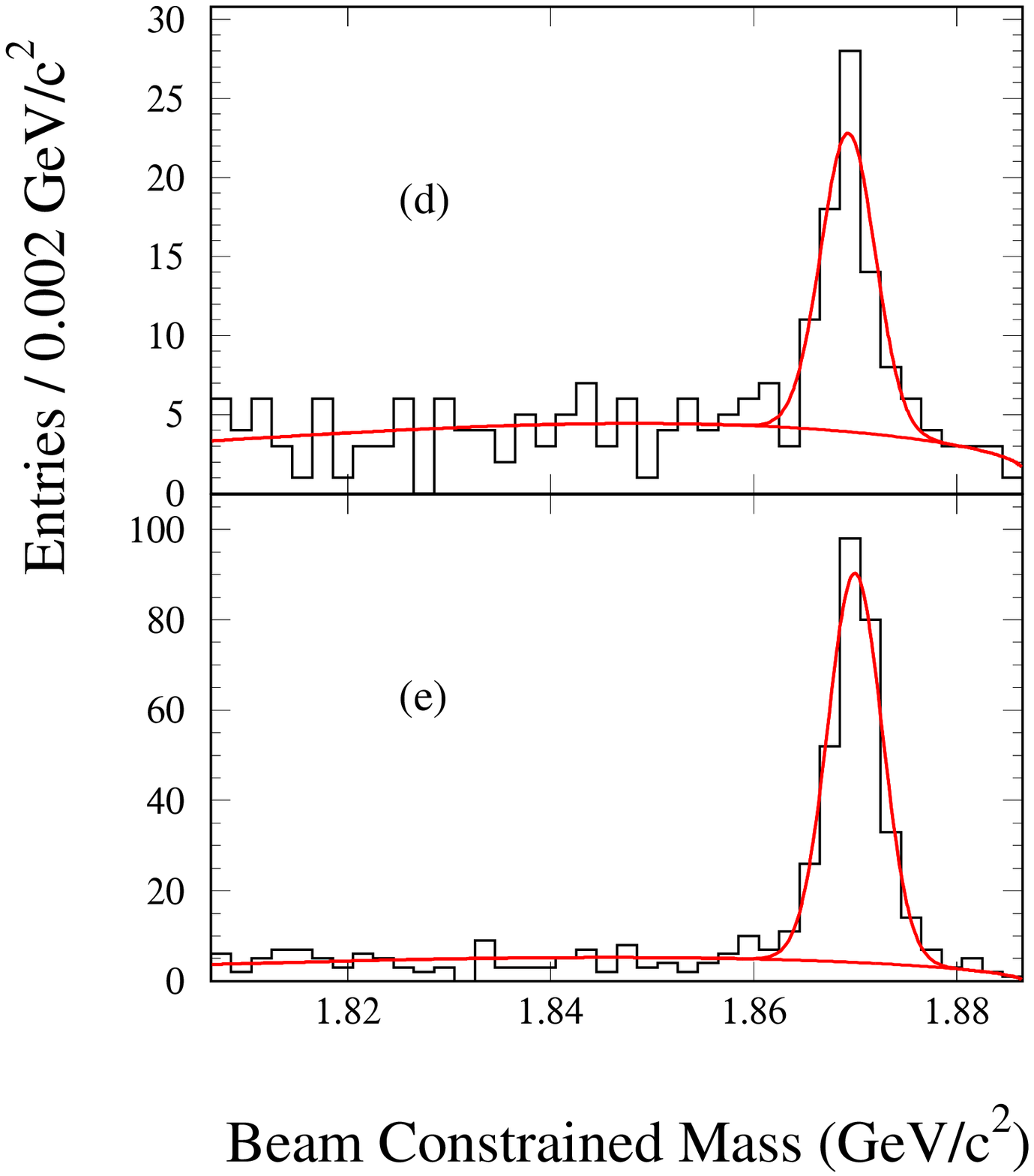}
  \caption{Beam energy constrain mass distributions for decays
   (a) $D^0 \rightarrow K^- K^+$, (b) $D^0 \rightarrow \pi^- \pi^+$,
   (c) $D^0 \rightarrow K^- \pi^+$ and
   (d) $D^+ \rightarrow \overline{K^0} K^+ $,
   (e) $D^+ \rightarrow \overline{K^0} \pi^+$.}
\label{2body}
\end{figure}
\begin{figure}[htbp]
  \centering
  \includegraphics[width=6.5cm]{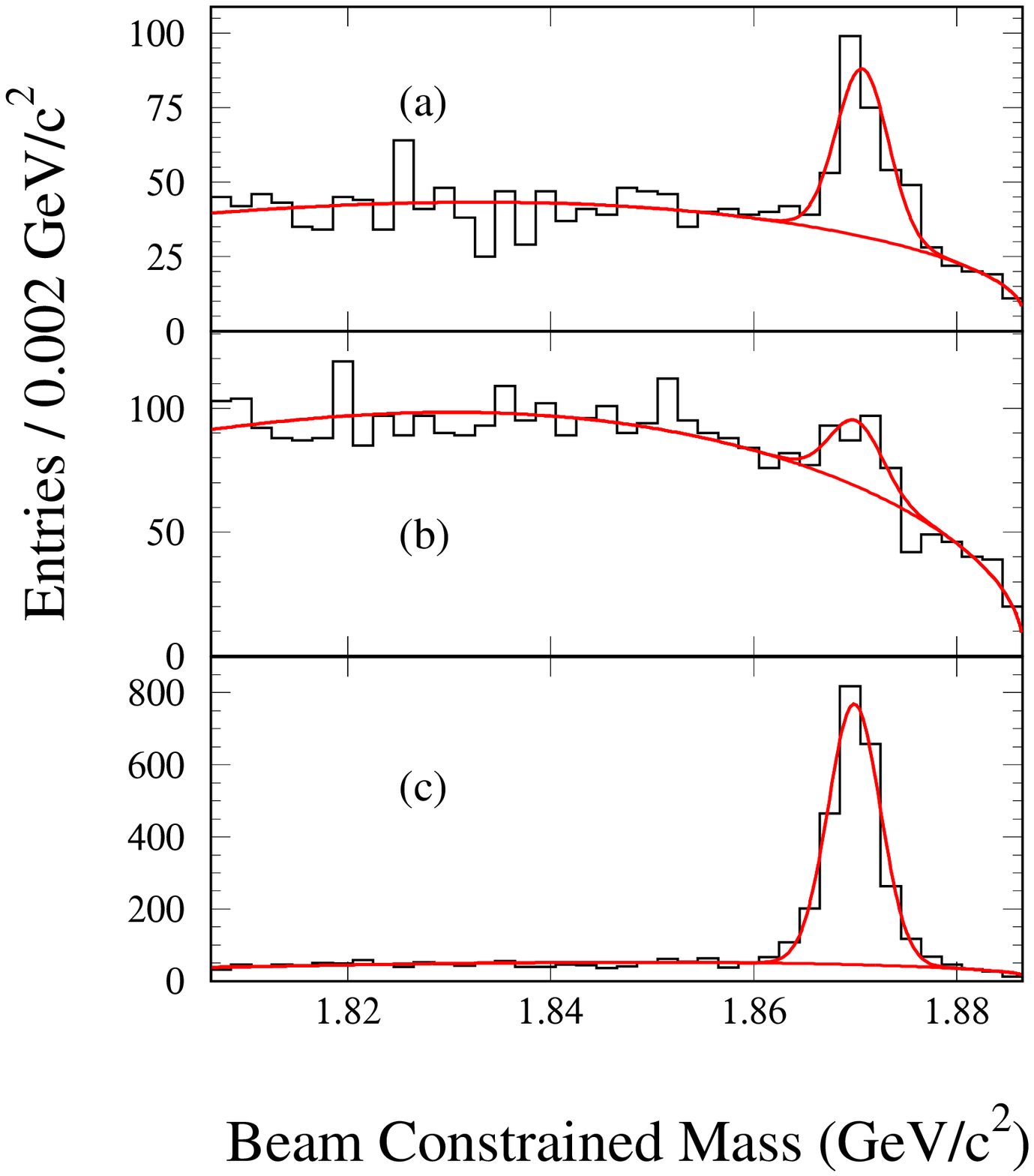}
  \includegraphics[width=6.5cm]{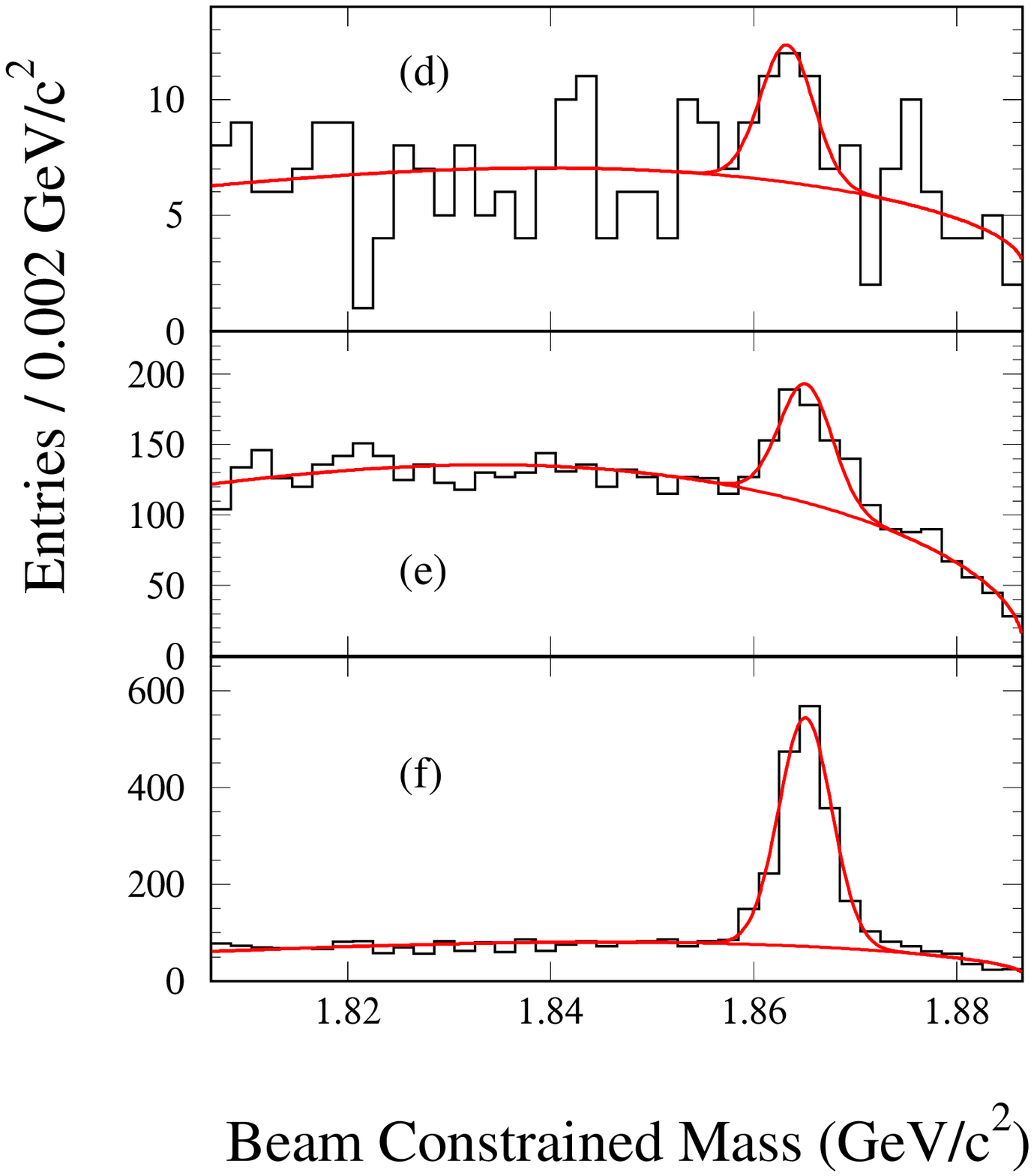}
  \caption{Beam energy constrain mass distribution for decay
  (a) $D^+ \rightarrow K^- K^+ \pi^+$, (b) $D^+ \rightarrow \pi^- \pi^+ \pi^+$,
  (c) $D^+ \rightarrow K^- \pi^+ \pi^+$ and
  (d) $D^0 \rightarrow K^- K^+ \pi^+ \pi^-$,
  (e) $D^0 \rightarrow \pi^- \pi^+ \pi^+ \pi^-$,
  (f) $D^0 \rightarrow K^- \pi^+ \pi^+ \pi^-$.}
\label{3body}
\end{figure}

One event could be counted more than
once as a tag candidate. In order to calculate the actual number
of tagged events in an unbiased manner, the following criterion is
applied to select only one tag combination per event: if more than
one combination of tracks form the desired tag, the combination is
chosen when the lowest momentum track of this combination has the
largest momentum of all other combinations. The result mass plots
are shown in Figure~\ref{2body}, ~\ref{3body}.\par

\section{DETECTOR ACCEPTANCE}
\label{secmc} The detection efficiency is determined by a detailed
Monte Carlo simulation of $D \overline{D}$ production, decay and
detector response. The decay branching ratios of neutral and
charged $D$ meson are taken from the world average values
\cite{PDG}, some unseen decay modes are set according to the rules
of isospin conservation.
 Simulated
events are processed through the event reconstruction, selection
and analysis program.\par There are several sub-resonant decay
modes in 3-body and 4-body Cabibbo suppressed channels. The
detection efficiency is not uniform among these decay modes. The
relative decay branching fractions and their errors in PDG are
quoted. For $D^{+}\rightarrow\pi^{-}\pi^{+}\pi^{+}$ channel,
$\rho^{0}\pi^{+}$ and $\pi^{-}\pi^{+}\pi^{+}$ modes are
considered; for $D^{+}\rightarrow K^{-} K^{+}\pi^{+}$ channel,
$\phi\pi^{+}$, $\overline{K^{*0}}K^{+}$ and $K^{-}K^{+}\pi^{+}$
are considered; for $D^{0}\rightarrow K^{-} K^{+}\pi^{+}\pi^{-}$
channel, $\phi\pi^{+}\pi^{-}$, $\phi\rho^{0}$,
$\overline{K^{*0}}K^{*0}$, $K^{-}K^{+}\rho^{0}$ and
$K^{-}K^{+}\pi^{+}\pi^{-}$ are considered; for
$D^{0}\rightarrow\pi^{-}\pi^{+}\pi^{+}\pi^{-}$ channel, 
the uncertainty to
Monte Carlo efficiency is estimated to be less than 2$\%$.
\par

\section{RESULTS}
\label{secbr} The observed number of each Cabibbo suppressed decay
channel is
determined by fitting the distributions to a function of the form:
\begin{equation}
\begin{array}{ccc}
F(m)&=&a_{\mathrm{1}}\left[m\sqrt{1-\left(\frac{m}{E_{\mathrm{b}}}\right)^{2}}
\exp\left(a_{\mathrm{2}}\left[1-\left(\frac{m}{E_{\mathrm{b}}}\right)^2\right]\right)\right]\\
 &+
&a_{\mathrm{3}}\exp\left(-\frac{(m-M_{\mathrm{D}})^{2}}{2\sigma^{2}}\right)
+ a_{\mathrm{4}}
\end{array}
\end{equation}
where the first term parameterizes the background; the Gaussian
term accounts for the signal. Just above the $D$ mass, there is no
more phase space available for a decay to a pair of
$D\overline{D}$ mesons. The background term is ARGUS form
\cite{argusbg}, where $a_{\mathrm{1}}$ is the normalization
factor; $a_{\mathrm{2}}$, a scale factor for the exponential term;
$E_{\mathrm{b}}$, the beam energy and fixed at 1.8865GeV while
fitting the mass plot. For $\psi(3770)$ scan data, the beam energy
is not a constant, but the background shape is similar, a constant
background term $a_{\mathrm{4}}$ is applied to evaluate the
varying beam energy points. The mass resolution of each Cabbibo
suppressed mode is determined by fitting the $M_{\mathrm{bc}}$
plot of similar Cabbibo allowed modes. The event number for each
mode is summarized in Table~\ref{results}. \par The pions that are
part of reconstructed $K^{0}_{S}$'s, are not used in modes
$\pi\pi$, $~\pi\pi\pi$, $~\pi\pi\pi\pi$ and $KK\pi$, $KK\pi\pi$.
However, there remain some $D$ decays into $K^{0}_{S}\pi^{+}$,
$K^{0}_{S}\pi^{+}\pi^{-}$ and $K^{-}K^{+}K^{0}_{S}$, where the
$K^{0}_{S}$ is not identified as a separated vertex, which will be
the major background source for decay mode
$D^{+}\rightarrow\pi^{-}\pi^{+}\pi^{+}$ and
$D^{0}\rightarrow\pi^{-}\pi^{+}\pi^{+}\pi^{-},
K^{-}K^{+}\pi^{+}\pi^{-}$. To reduce the feed-down
$K^{0}_{S}\rightarrow\pi^{+}\pi^{-}$ background, a cut of
$\left|M_{\pi^{+}\pi^{-}}-M_{K^{0}}\right|>0.040$ GeV$/c^{2}$ is
imposed on the invariant mass of each pion pairs for
$\pi^{-}\pi^{+}\pi^{+}$ and $K^{-}K^{+}\pi^{+}\pi^{-}$ modes.\par
\begin{figure}[htbp]
\centering
\includegraphics[width=6.0cm]{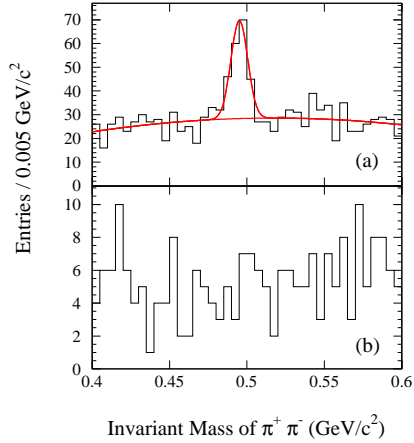}
\caption{ $M_{\pi^{+}\pi^{-}}$ for
$\pi^- \pi^+ \pi^+ \pi^-$ candidates within $D^0$ signal region (a) and
sideband region (b).}
\label{4piks}
\end{figure}

For the decay $D^{0}\rightarrow\pi^{-}\pi^{+}\pi^{+}\pi^{-}$,
plots of the invariant mass of all $\pi^{+}\pi^{-}$ combinations
in $D^0$ candidates within the signal and sideband regions are
shown in Figure~\ref{4piks}(a) and (b) respectively. There is a
clear $K^{0}_{S}$ peak within $D^0$ signal region with a fitted
events number of $112.5 \pm 19.0$ events and no clear $K^0_S$ peak
within the sideband region. The $K^0_S$ number is consistent with
the expected background (Monte Carlo study gives this number as
98.8) due to $D^{0}\rightarrow\overline{K^{0}}\pi^{+}\pi^{-}$.
These are thus subtracted from the $4\pi$ signal.\par

\section{SYSTEMATIC UNCERTAINTIES}
\label{secse} In this analysis, we normalize the relative ratios
of Cabibbo suppressed decay to similar Cabibbo favoured modes,
which permits the cancellation of many common systematic errors.
The systematic uncertainties on energy cut and particle
identification can not be canceled completely.\par Systematic
uncertainties on particle identification and $\Delta E$ cuts are
about 1-3 \% and 1-6\% contributing to the relative fractions
respectively.
 The systematic
uncertainties on background subtraction due to the $K^{0}_{S}$
contamination are estimated to be $\sim2\%$ \cite{ks}.\par The
detection efficiencies are not uniform among the different
sub-resonant decay modes in the 3 and 4 body decays, $0.5-2.0\%$
are estimated for different modes $\pi^{-}\pi^{+}\pi^{+}$,
$K^{-}K^{+}\pi^{+}$ and $\pi^{-}\pi^{+}\pi^{+}\pi^{-}$. Large
uncertainty in $K^{-}K^{+}\pi^{+}\pi^{-}$ mode is found due to the
decays of low momentum kaons.

\section{CONCLUSIONS}
\label{secconcl}

\begin{table}[h]
\caption{Measurement results of 7 Cabibbo suppressed decay
modes.}
\label{results}

\begin{center}

\begin{tabular}{cccc}
\hline \hline

Decay mode & Yield & Relative efficiency & Branching ratio \\ \hline

{\large $\frac{K^- K^+}{K^- \pi^+}$} &
{\large $\frac{242.2 \pm 20.1}{1934 \pm 49}$} &
$1.029\pm0.017$ & $0.122 \pm 0.011 \pm 0.004$ \\[1.0ex]

{\large $\frac{\pi^- \pi^+}{K^- \pi^+}$} &
{\large $\frac{75.9 \pm 14.7}{1934 \pm 49}$} &
$1.146\pm0.030$ & $0.034 \pm 0.007 \pm 0.001$ \\[1.0ex]

{\large $\frac{\overline{K^0} K^+}{\overline{K^0} \pi^+}$} &
{\large $\frac{63.2\pm9.8}{287 \pm 18}$} & $0.991\pm0.041$ &
$0.222 \pm 0.037 \pm 0.013$ \\[1.0ex]

{\large $\frac{K^- K^+ \pi^+}{K^- \pi^+ \pi^+}$} &
{\large $\frac{181.2\pm20.2}{2324 \pm 53}$} &
$0.669 \pm 0.015 \pm 0.010$ & $0.117 \pm 0.013 \pm 0.007 $ \\[1.0ex]

{\large $\frac{\pi^- \pi^+ \pi^+}{K^- \pi^+ \pi^+}$} &
{\large $\frac{84.9\pm22.4}{2324 \pm 53}$} &
$0.888 \pm 0.029 \pm 0.004$ & $0.041 \pm 0.011 \pm 0.003 $ \\[1.0ex]

{\large $\frac{K^- K^+ \pi^+ \pi^-}{K^- \pi^+ \pi^+ \pi^-}$} &
{\large $\frac{19.3\pm8.0}{1540 \pm 51}$} & $0.286 \pm 0.021 \pm
0.017$ &
$0.044 \pm 0.018 \pm 0.005$ \\[1.0ex]

{\large $\frac{\pi^- \pi^+ \pi^+ \pi^-}{K^- \pi^+ \pi^+ \pi^-}$} &
{\large $\frac{(274.4\pm31.8)-(112.5 \pm 19.0)}{1540 \pm 51}$} &
                                                $1.336 \pm 0.028 \pm 0.027$ &
$0.079 \pm 0.018 \pm 0.005 $ \\[1.0ex]

\hline \hline

\end{tabular}
\vskip 1.5cm
\end{center}
\end{table}

The relative fractions of seven Cabibbo suppressed decay modes are
tabulated in Table.\ref{results}. For each mode, fitted event
number, background number, relative efficiency and relative
branching ratio are listed. The errors of relative efficiency are
Monte Carlo statistical error and systematic error due to
sub-resonant modes respectively. The first error of branching
ratio is statistical, the second one is systematic. The estimated
systematic uncertainty is 3-6\% for all modes, and the statistical
uncertainties of the measurements are about 10\% or greater.
Results from this measurement are consistent with the world
average values and we have improved the previous measurements of
the $D^+ \rightarrow \overline{K^0} K^+$ relative branching ratio.
The measurements of Cabibbo suppressed branching ratios presented
here provide new insights into the mechanism of nonleptoinc $D$
decays. Exact SU(3) symmetry predicts the equality of $\Gamma (D^0
\rightarrow \pi^- \pi^+)$ and $\Gamma (D^0
\rightarrow K^- K^+)$. 
But the above results show they are not equal. Several distinct
effects could contribute to this inequality. Final states
interactions breaking SU(3) symmetry could however account for the
difference.

\section{ACKNOWLEDGMENTS}
\label{secack}
   The BES collaboration thanks the staff of BEPC for their hard efforts.
This work is supported in part by the National Natural Science
Foundation of China under contracts Nos. 19991480, 10225524,
10225525, the Chinese Academy of Sciences under contract No. KJ
95T-03, the 100 Talents Program of CAS under Contract Nos. U-11,
U-24, U-25, and the Knowledge Innovation Project of CAS under
Contract Nos. U-602, U-34 (IHEP); and by the National Natural
Science Foundation of China under Contract No. 10175060 (USTC),
and No. 10225522 (Tsinghua University).

\end{document}